\documentclass{article}
\usepackage{caption}
\usepackage{subcaption}
\usepackage{svg}
\usepackage{PRIMEarxiv}
\usepackage{cite}
\usepackage{amsmath,amssymb,amsfonts}
\usepackage{algorithmic}
\usepackage{graphicx}
\usepackage{siunitx}
\usepackage{textcomp}
\usepackage{upgreek}
\usepackage[utf8]{inputenc} % allow utf-8 input
\usepackage[T1]{fontenc}    % use 8-bit T1 fonts
\usepackage{hyperref}       % hyperlinks
\usepackage{url}            % simple URL typesetting
\usepackage{booktabs}       % professional-quality tables
\usepackage{amsfonts}       % blackboard math symbols
\usepackage{nicefrac}       % compact symbols for 1/2, etc.
\usepackage{microtype}      % microtypography
\usepackage{lipsum}
\usepackage{fancyhdr}       % header
\usepackage{graphicx}       % graphics
\graphicspath{{media/}}     % organise your images and other figures under media/ folder

\title{Estimations of lung structural properties from a single propagation-based dark-field X-ray image
}
\author{
  D. W. O'Connell\textsuperscript{1}, K. S. Morgan\textsuperscript{1}, L. C. P. Croton\textsuperscript{1}, J. A. Pollock\textsuperscript{1}, G. Ruben\textsuperscript{2}\\
  \textbf{K. J. Crossley\textsuperscript{3,4}, M. J. Wallace\textsuperscript{3,4}, 
  E. V. McGillick\textsuperscript{3,4},  S. B. Hooper\textsuperscript{3,4}}\\
  \textbf{ and M. J. Kitchen\textsuperscript{1,3}}\\
\textsuperscript{1}
School of Physics and
Astronomy, Monash University, Clayton, VIC 3800, Australia\\
\textsuperscript{2}
Australian Synchrotron, ANSTO, Clayton, VIC 3168, Australia\\
\textsuperscript{3} The Ritchie Centre, Hudson Institute of Medical Research\\
\textsuperscript{4} Monash Department of Obstetrics and Gynaecology,\\ Monash University, Clayton, VIC 3800, Australia\\
\texttt{email: dylan.oconnell1@monash.edu}}

\begin{document}
\maketitle

\begin{abstract}
In this investigation, we applied a single-projection dark-field imaging technique to gain statistical information on the smallest airway structures within the lung---the alveoli---focusing on their size and number as key indicators of lung health. The algorithm employed here retrieves the projected thickness of the sample from a propagation-based phase contrast image using the transport-of-intensity equation. The first Born approximation is then used to isolate the dark-field signal associated with edge scattering, which increases the visibility of microstructure boundaries. PMMA spheres of known sizes were imaged first as an idealised alveolar model. The dark-field signal was then recovered from propagation-based phase-contrast X-ray images of the lungs of small mammals using this method. The retrieved dark-field signal was found to be proportional to both the alveolar size ($R^2 = 0.85$) and the number in projection ($R^2 = 0.69$), and these measurements could be combined to provide an estimate of the total surface area of the alveolar interfaces ($R^2 = 0.78$). This demonstrates the approach's ability to indicate lung health using the dark-field signal retrieved from a single-phase-contrast X-ray image. 

\end{abstract}

\keywords{dark-field, dynamic, phase contrast, lung, X-ray imaging}

\section{Introduction}
\label{sec:introduction}
Chronic obstructive pulmonary disease (COPD) has one of the highest rates of morbidity and mortality worldwide~\cite{vestbo_global_2013,adeloye2022global}. Despite this, diagnosis often occurs in later stages, by which point global respiratory function has declined significantly~\cite{csikesz_new_2014,fazleen2020early}. This is because current diagnostic techniques typically measure the volume and airflow rate of the whole lung. Imaging techniques such as magnetic resonance imaging (MRI), sonography, and X-ray imaging can be used to assess regional lung changes. The spatial resolution of both sonography and MRI is on the order of $>1$~mm, which is much larger than the distal airway structures within the lung, such as alveoli, which are on the order of 100 microns~\cite{foster2014mouse}. The ability to retrieve information about the alveoli, such as their size and number, can provide vital information about lung diseases such as COPD~\cite{milne2014advanced, kitchen_emphysema_2020}. Custom high-resolution research-focused CT imaging is the only method to directly and non-invasively measure the alveolar dimensions. However, since micro-CT requires many projection images (typically, hundreds to thousands), it has a much higher associated radiation dose than a single projection image, as dose is proportional to the number of images. Lung CT also relies on breath control or respiratory-gated image acquisition to avoid motion artifacts~\cite{ dubsky_synchrotron-based_2012, pialat2012visual} and suffers from relatively poor temporal resolution~\cite{guerrero_dynamic_2006, pan_4d-ct_2004}. These limitations of conventional lung imaging motivates the development of a new approach to obtain information about the lungs at the alveolar level from a single projection X-ray image.  

Alveoli within the lung act as scattering objects and produce ultra-small-angle X-ray scattering (USAXS)~\cite{pelliccia2017theory}, a main contributor to what is known as X-ray dark-field imaging~\cite{kitchen_emphysema_2020, gradl2018dynamic, paganin2019x}. This is especially important in the context of COPD, which is characterised by airway inflammation resulting in thickening and narrowing of the airways~\cite{adeloye2022global}.

Dark-field images can be measured using several X-ray imaging techniques~\cite{esposito2023laboratory}. Key examples include analyser-based imaging  (ABI)~\cite{kitchen_emphysema_2020}, Edge illumination~\cite{Endrizzi2015}, speckle-based imaging~\cite{zdora2017x} and grating interferometry (GI)~\cite{gradl2018dynamic, rauch2020discrimination}. Using ABI, the number of scattering objects can be recovered by measuring the rocking curve parameters recorded with and without the sample~\cite{kitchen_emphysema_2020}. In GI, the visibility reduction of the interferogram reference pattern is similarly related to the size of the scatterers~\cite{taphorn_grating-based_2020,willer2021x}. 

At high imaging resolutions, arbitrary reference patterns such as sandpaper~\cite{morgan2012x} or other irregular surfaces~\cite{zdora2017x} can be used to produce a near-field speckle pattern, which will reduce in visibility in the presence of dark-field signal ~\cite{berujon2012x, zanette2014speckle, alloo2023m}. Such an approach, simply using a sieve instead of sandpaper, has recently been applied to image the lungs of small animals in motion~\cite{how2024vivo}. Other recent techniques have used the sample as its own reference pattern to extract a dark-field image from multiple sample-to-detector distances~\cite{leatham2023x} or energies~\cite{ahlers2024x} in a propagation-based setup~\cite{cloetens1996phase,snigirev1995possibilities}. 

Propagation-based imaging (PBI) methods have previously been used to quantify alveolar size in the lungs without extracting a dark-field signal, but instead by studying speckle intensity patterns in the image plane caused by the lung~\cite{kitchen2004origin}. In lung imaging, the speckle pattern is produced when the alveoli refract the X-ray beam, with the properties of the frequency space spectrum---such as the dominant frequency and the total power of the spectrum---shown to be related to both the number and dominant size of the alveoli~\cite{leong_measurement_2013,leong_real-time_2014,kitchen_x_ray_2015}. The ability to retrieve information without using a reference pattern and within a single exposure is advantageous, as it allows for application to dynamic systems. For lung imaging, the ability to characterise the structures of the airways can provide vital information about lung health~\cite{hooper2009imaging}. 

The X-ray phase information contained in a single propagation-based image can also be extracted using phase-retrieval algorithms. The most widely adopted phase retrieval algorithms in PBI have been based on the transport of intensity equation~\cite{teague1983deterministic} and assume a homogeneous material. This allows phase effects in PBI (often exhibited as edge enhancement) to be recovered~\cite{paganin_simultaneous_2002}, achieving a significant boost in signal-to-noise ratio (SNR)~\cite{kitchen2017ct, croton2018situ} from a single exposure. Recently, Gureyev et al. developed an algorithm to retrieve dark-field images from PBI that can improve visibility of microcalcifications in breast tissue~\cite{gureyev_dark-field_2020, aminzadeh2022imaging}. This was done by retrieving a signal that shows where diffracted wave components are present in the object plane to obtain information about the small features within the image (see Section~\ref{sec:alg}).

Starting with the assumption that scattering objects refract incoming radiation as it traverses a material, it has been theorised that, for relatively large microstructures, the strength of the USAXS signal should scale with the number of interfaces while remaining relatively independent of their size~\cite{von1926refraction}. Indeed, this behaviour has been demonstrated for analyser-based imaging (ABI)~\cite{kitchen_emphysema_2020} and single-grid imaging~\cite{how2022quantifying}. However, dark-field extraction from GI has been shown to correlate with both microstructure size and sample thickness~\cite{taphorn_grating-based_2020, vignero2018translation, gassert2021x, urban2022qualitative}. This raises the question of which microstructure properties, namely size and number, the algorithm in this investigation will be derived from. The following investigation uses an algorithm which explicitly measures intensity variations due to diffraction and thus depends on the number and size of the structures generating phase contrast rather than solely on accumulated scattering interfaces. Consequently, we are not measuring exactly the same phenomenon as ABI or GI, which often emphasises multiple scattering events (USAXS), leading to reduced visibility or diffracted beam intensity. The dark-field technique of Gureyev et al. may therefore provide distinct, and potentially complementary, information about the sample microstructure.

Herein, we aimed to assess (1) the viability of the algorithm of Gureyev et al.~\cite{gureyev_dark-field_2020} to retrieve dark-field signals from a single PBI projection to estimate microstructure size and number and (2) its use for real-time imaging during respiration, where the microstructure is alveoli. 
 
\section{Image Reconstruction Methods}
\label{sec:alg}

The dark-field algorithm developed by Gureyev et al.~\cite{gureyev_dark-field_2020} begins by separating an image $I_{(R,k_x,k_y)}$ into low- and high-diffractive orders within the image. Here, $R$ denotes the effective propagation distance from the object to the image plane, and $(k_x,k_y)$ are the coordinates in Fourier space.  The remaining low diffracted order we will label as the bright-field signal, such that we can write:
\begin{equation}
    I_{(R,k_x,k_y)} = I_{(R,k_x,k_y)}^{bright} + I_{(R,k_x,k_y)}^{dark}.
\end{equation}
 $I_{(R,k_x,k_y)}^{bright}$ describes the slowly-varying signal (i.e. low diffractive orders) that can be resolved on the detector and is the larger of the two terms. The bright-field term can be expressed by the transport-of-intensity equation (TIE) approximation. Phase retrieval algorithms such as Paganin's algorithm~\cite{paganin_simultaneous_2002, beltran_2d_2010} can propagate backward to the object plane using the TIE. This is a common approach in phase contrast imaging that uses both refraction and attenuation information to retrieve the sample projected thickness information; a full description of the mathematics can be found in \cite{paganin_simultaneous_2002}). 

$I_{(R,k_x,k_y)}^{dark}$ is a weaker and more rapidly varying signal~\cite{gureyev_dark-field_2020, gureyev2004optical}. A major contributing factor to this signal is the USAXS fans caused by scattering from interfaces/edges. In this way, an unresolvable structure can give measurable intensity fluctuations, which can be represented by the first Born approximation~\cite{paganin2006coherent}. 

For a single material. we can express the bright-field component in the object plane as~\cite{paganin_simultaneous_2002}
\begin{equation}
\label{tie_ret}
    I_{(0,k_x,k_y)}^{TIE} = \frac{I_{(R,k_x,k_y)}}{1 + \gamma_1 \pi \lambda R (k_x^2+k_y^2)}.
\end{equation}
Here, \(\lambda\) is the wavelength and $\gamma_1$ is the ratio of the imaginary and real parts of the complex refractive index ($\gamma = \delta / \beta$), $n=1-\delta + i\beta$. $I_{(0,k_x,k_y)}^{TIE}$ can then be numerically free-space-propagated to the image plane to give $I_{(R,k_x,k_y)}^{TIE}$. The dark-field signal in the image plane can then be calculated as $I_{(R,k_x,k_y)}^{dark} = I_{(R,k_x,k_y)} - I_{(R,k_x,k_y)}^{TIE}$. To compute the distribution of the sources of the dark-field signal in the object plane, we can evaluate it as~\cite{gureyev_dark-field_2020}:
\begin{equation}
\label{ret}
I_{(0,k_x,k_y)}^{dark} = \frac{I_{(R,k_x,k_y)}^{dark}}{2 \cos[\pi \lambda R(k_x^2+k_y^2)] + 2\gamma_2 \sin[\pi \lambda R(k_x^2+k_y^2)}.
\end{equation}
This can be translated into an iterable version of $m$ iterations as~\cite{aminzadeh2022imaging}:
\begin{equation}
I^{dark, m}_{(R,k_x,k_y)} = I_{(R,k_x,k_y)} - I^{TIE}_{(R,k_x,k_y)} - I^{dark, m-1}_{(R,k_x,k_y)}.
\label{eq:dark_field_iteration}
\end{equation}
$I^{dark, m-1}_{(R,k_x,k_y)}$ is calculated via numerical free-space
forward propagation of the homogeneous complex amplitude $I^{(dark, m-1)}_{(0,k_x,k_y)}$ with the phase, $\varphi$ as $\varphi^{(dark, m-1)}_{(0,k_x,k_y)} = 0.5 \gamma_2\ln\left[I^{(dark, m-1)}_{(0,k_x,k_y)}\right]$. The iterable version of this algorithm is available on GitHub~\cite{github}. This algorithm has been shown to provide better image quality (such as object visibility and contrast-to-noise ratio). Previously~\cite{aminzadeh2022imaging}, this algorithm has been shown to behave in a semi-convergent manner. To know how many iterations are required, we set an end criterion where the increase in the root mean square (RMS) norm between subsequent iterations $I_{(R,k_x,k_y)}^{dark, m}$ ($m$ = 1,2,3...) is less than the RMS error of the original image. For further details, see~\cite{aminzadeh2022imaging}. Additionally, when applying the algorithm, $\gamma$ is free to have different values between Equations~\ref{tie_ret} and~\ref{ret}, which is relevant for multi-material samples where the slowly varying component material may be used as the first $\gamma$ value `$\gamma_1$', and for the material that causes the rapidly varying signal we can apply `$\gamma_2$' for the retrieval step in Equation~\ref{ret}.

It should be noted that the above discussion of the dark-field signal primarily relates to the measurable intensity variations resulting from scattering from the edges~\cite{paganin2006coherentborn} of the unresolved microstructure within the sample, where other dark-field methods discussed in Section~\ref{sec:introduction} focus on the reduction in visibility of reference patterns from interferometric methods. This observation implies that the measured signal will be highly sensitive to the boundaries between media, particularly where the assumption of homogeneity breaks down. The resultant signal will also have both positive and negative components for which we subtract 1 from the distribution to center it at zero, then take the the moduli to represent the dark-field component and (see Figure~\ref{fig:stepwedge}(d) for a visual example).

\section{Animal Handling and Ethics}

This experiment used rabbit kittens that had been used in experiments conducted with approvals from the Australian Synchrotron and Monash University Animal Ethics Committees. All experiments were performed in accordance with relevant guidelines and regulations. The kittens were humanely killed in line with approved guidelines and the carcasses scavenged for this experiment.

\section{Experimental Setup and Imaging Procedure}
\label{sec:experiment}

\begin{figure}[t!]
\centering
\includegraphics{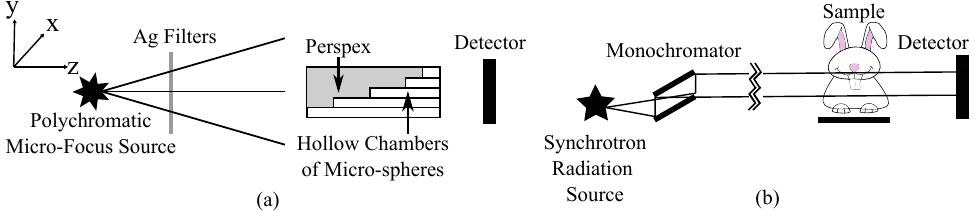}
\caption{(a) The experimental setup using a polychromatic micro-focus X-ray source, filtered with 51~$\mathrm{\upmu m}$ of silver (Ag)  to give a quasi-monochromatic spectrum with a mean energy of 21~keV at 50~kVp. An object-to-detector distance of 0.78~m was used to obtain PBI projections of the step wedge filled with PMMA spheres of a given size range. (b) The arrangement of optical elements at the Imaging and Medical Beamline (IMBL) of the Australian Synchrotron~\cite{stevenson2017quantitative}. Here, an X-ray source-to-sample distance of 140~m was used with a sample-to-detector distance of 1.5~m. A double Laue bent-crystal monochromator obtained monochromatic X-rays from the source radiation at 25~keV.  Rabbit kittens that had been humanely killed were taped securely across their forelegs and upper chest to a perspex holder that was affixed to a rotation stage to obtain full-field CTs.}
\label{fig:experiment}
\end{figure}

To explore how the dark-field reconstruction outlined in Section~\ref{sec:alg}  relates to the size and number of scattering objects, we initially imaged Poly(methyl methacrylate) (PMMA) spheres (\textit{Cospheric, LLC}, USA)  contained within a hollow step wedge with adjacent chambers of approximate depth of  1.5, 2.5, 5.5, 10.6, 15.8, and 20.8~mm $\pm$ 0.5~mm. We used five sets of spheres, where each set was sieved between two size limits, with the specifications provided in Table~1.

\begin{table}[h!]
\label{tab:spheres}
\centering
\begin{tabular}{ |p{2cm}||p{2cm}|p{2cm}|  }
 \hline
 \multicolumn{3}{|c|}{Poly(methyl methacrylate) sphere sizes} \\
 \hline
Minimum size ($\mathrm{\upmu m}$)& Maximum size ($\mathrm{\upmu m}$)& Median size ($\mathrm{\upmu m}$)\\
 \hline
 90 &   106  & 98  \\
  125 &150 & 137.5\\
   150    &180 & 165\\
 212&   250  & 231\\
 250  & 300    &275\\
 \hline
\end{tabular}
\caption{Sieved PMMA sphere size properties.}
\end{table}

For all particle sizes, we recorded a PBI image (see Figure~\ref{fig:experiment}(a)) of the filled step wedge to obtain measurements of varying thickness (or, equivalently, particle number). This was performed using a micro-focus X-ray source (THE-Plus, \textit{X-RAY WorX} GmbH, Germany) at the $\delta$  \textit{X-ray Laboratory} at Monash University. The source had a silver transmission target with the tube voltage and target power set to 50 kV and 15 W, respectively. The beam was filtered with 51~$\mathrm{\upmu m}$ of silver and had a mean energy of around 21~keV. The images were recorded with a Medipix3 photon-counting detector (WidePIX L 2$\times$5, \textit{Advacam}, Czech Republic) set to charge-summing mode, where X-ray photons measured to have been below 3.9~keV were ignored on an individual pixel basis, and clusters that are integrated to be below 16~keV were ignored. An effective pixel size of 7~$\mathrm{\upmu m}$ was used with an exposure time of 90 seconds. Subsequently, the dark-field signal was retrieved from each flat-field-corrected PBI image, using the method described in Section~\ref{sec:alg} with $\gamma_1 = \gamma_2 = 2{,}281$ ( set for PMMA spheres in air with 21~keV X-rays).  

\begin{figure}[b!]
\centering
\includegraphics{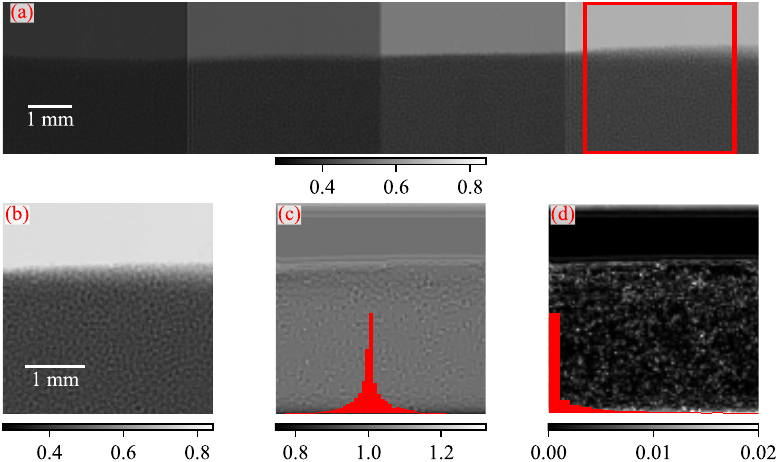}

\caption{(a) Propagation-based phase contrast image of the 5.5-20.8~mm chambers of a step wedge filled with the 125-150~$\mathrm{\upmu}$m spheres. The red box in (a) is magnified in (b). After applying the dark-field retrieval algorithm, we obtain (c), where a histogram of the returned values is overlaid in red. Following the explanations in Section~\ref{sec:alg}, we take the modulus of this image and obtain a better visual representation of the dark-field in (d).}
\label{fig:stepwedge}
\end{figure}
We also investigated whether a dark-field signal produced by the alveoli in the lungs can be recovered following the image reconstruction method described in Section~\ref{sec:alg}. To do this, we imaged five New Zealand white rabbit kittens delivered by caesarean. We performed high-resolution breath-hold CTs (fixed pressure of nitrogen $\approx$~40cmH$_2$O) at the Imaging and Medical Beamline (IMBL) of the Australian Synchrotron~\cite{stevenson2017quantitative} (see Figure~\ref{fig:experiment}(b)) to map the size and number of alveoli throughout the lung volume. We recorded 1800 projections over 180 degrees at a 25~keV X-ray energy. The projections were recorded with an exposure time of 30~ms per projection at an object-to-detector propagation distance of 1.5~m. They were recorded with a lens-coupled sCMOS detector (pco.edge 5.5, \textit{Excelitas Technologies Corp}, USA) with an effective pixel size of 14.9~$\mathrm{\upmu m}$ using a 25~$\mathrm{\upmu m}$  thick gadolinium oxysulfide (Gadox, Gd$_2$O$_2$S:Tb) phosphor. To demonstrate the applicability for dynamic imaging, images were also acquired during mechanical ventilation sequences, with seven images acquired per 1.5 second breath sequence (between 5 and 40~cm H$_2$O). Each frame had an exposure time of 15~ms. Each animal was fixed in the anterior-posterior orientation during the ventilation sequence. Each PBI image was then processed to obtain the dark-field signal with $\gamma_1$ set for that of water in adipose tissue (387.23), as this is the slowly varying component of the image , and $\gamma_2$ that of air embedded in adipose tissue (2401.39), the more rapidly varying component.

\section{Dark-field Retrieval from PMMA Microspheres}
\label{sec:spheres}
\begin{figure}[t!]
\centering
\includegraphics{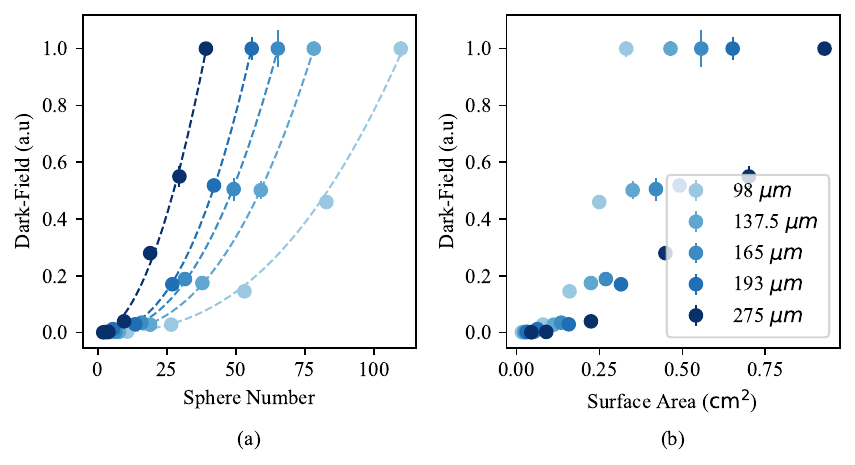}
\caption{Dark-field signal recovered for PMMA microspheres of different sizes plotted against the projected number of particles is shown in (a). In (b) we plot the dark-field signal against the micro-structure surface area. In both plots, each trend is normalised to the maximum dark-field in the trend.}
\label{fig:spherethicc}
\end{figure}
\begin{figure}[b!]
\centering
\includegraphics{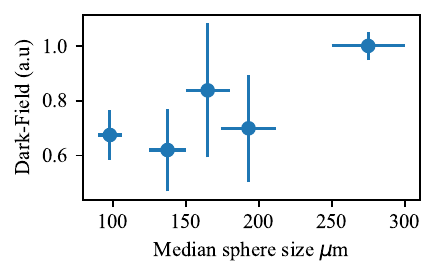}
\caption{For a given step-wedge thickness (15.6 mm here), we can plot the dark-field signal as a function of sphere size and observe a correlation ($R^2 = 0.69$) between scatterer size and dark-field signal.}
\label{fig:sphererad}
\end{figure}
Initial experiments were performed on five PMMA sphere sizes using the setup shown in Figure~\ref{fig:experiment}(a) to obtain PBI images for the step wedge filled with each sphere size. We applied the dark field retrieval algorithm described in Section~\ref{sec:alg} for each image. Figure~\ref{fig:stepwedge} compares the initial PBI image of the step wedge filled with one sphere size in (a,b) with the modulus of the dark-field reconstruction (subtracted by 1; see Section~\ref{sec:alg}) in (d). Note that the detector field of view was not large enough to capture the entire step wedge, so these images are stitched together. Figure~\ref{fig:spherethicc}(a) shows the the retrieved dark-field signal as a function of the projected number of particles seen in each pixel, calculated approximately by~\cite{kitchen_emphysema_2020}:
\begin{equation}
\label{eq:number_form}
    \mathrm{N} = \frac{\Gamma_{spheres} T_{step}}{2 r_{sphere}},
\end{equation}
where $\Gamma_{spheres}$ was the packing factor (calculated via fraction of attenuation) for PMMA spheres poured into the step wedge (0.54). Here, we can see that the dark-field signal increases as a function of particle number. Each particle size gives a different trend. The most applicable function for observed trends was a polynomial fit, $\propto N^n$, where $n$ ranges from 2 to 3, with these fits shown in Figure~\ref{fig:spherethicc}(a). 

The net surface area is an important structural parameter of particulate or porous matter. As discussed in Section~\ref{sec:alg}, this dark-field signal should be most sensitive to the unresolved interfaces of the microstructure. The microstructure surface area within a rectangular section of the sample can be calculated as $\mathrm{SA} = 4 \mathrm{N}\pi r^2$. This gives a single measurement to compare against the dark-field as we change the particle size or number. This trend is shown in Figure~\ref{fig:spherethicc}(b), where we see consistent trends for each sphere size. 

For a given step wedge thickness, we can also plot how the signal changes with increasing radius of the sphere $r_{sphere}$. Figure~\ref{fig:sphererad} shows the trend for a step wedge thickness of 15.6~mm. Here we observe a consistent positive trend, indicating that the dark-field signal is affected by both the particle number and size. The two data points with the weakest dark-field signal within each set, corresponding to the smallest step wedge thicknesses (1.5 and 2.5~mm) which had a low number of microstructures in projection and high attenuation, resulted in a negligible SNR. 

Dark-field retrieval performed better (i.e. showed higher correlation statistics; not shown) between the dark-field signal and changes in sphere size after the PBI images were binned using a 2$\times$2 kernel before reconstruction. A potential reason for this is that the edges of the spheres were too well resolved (original pixel size 7~$\mathrm{\upmu}$m) in the unbinned data, reducing the algorithm’s effectiveness in separating the scattering from the edges, as these effects mostly relate to the highest diffractive orders of the image. Alternatively, the cause could be that the near-field TIE approximation is a more accurate model for the near-field phase contrast effects at the binned pixel size, characterised by a larger Fresnel number.

\begin{figure}[t!]
\centering
\includegraphics{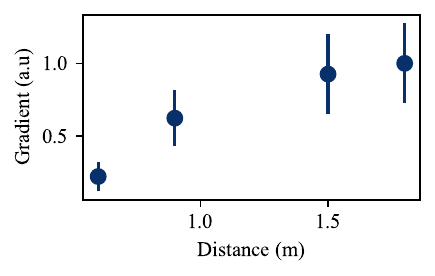}
\caption{At the setup shown in Figure~\ref{fig:experiment}(b), we vary the propagation distance from 0.6 to 1.8 m. For each propagation distance we measure the dark-field as a function of sphere size, averaged over the three largest step-wedge thicknesses. For each trend we calculate the gradient and plot this as a function of propagation distance.}
\label{fig:spheregrad}
\end{figure}  
\begin{figure}[b!]
\centering
\includegraphics{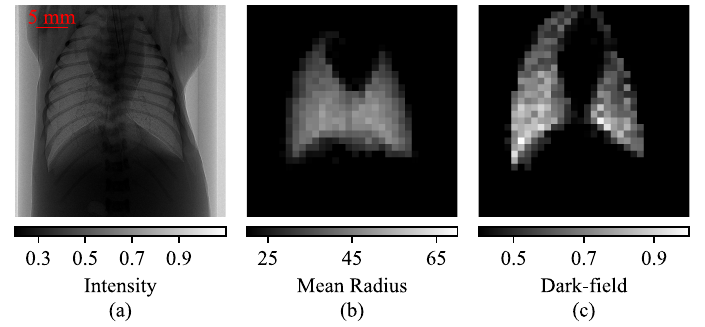}
        \caption{(a) shows the PBI projection of a rabbit kitten thorax in the anterior-posterior orientation. Following the granulometry steps outlined in Section~\ref{sec:lungs}, (b) shows the median alveoli radius projected along the optical axis in each binning volume. In the same binning region we can take the mean dark-field signal retrieved from a single PBI projection (c).}
        \label{fig:dfr}
\end{figure}

\section{Dark-field retrieval from Propagation-Based Lung Images}
\label{sec:lungs}

CT and dynamic PBI lung image sequences were acquired following the steps described in Section~\ref{sec:experiment} at the IMBL (Australian Synchrotron). As the dark-field retrieval algorithm requires phase-contrast effects, and these are heavily influenced by the object-to-detector distance, we sought to determine the optimum sample-to-detector propagation distance for the dark-field reconstruction approach. For this optimisation, we calculated the gradient of dark-field signal as a function of sphere size for each propagation distance. Images were captured over four object-to-detector distances between 0.6 and 1.8 m, the distance range previously determined for high speckle visibility at this beamline~\cite{o2022accurate}. The results for this optimisation are shown. In Figure~\ref{fig:spheregrad} the dark field signal appears to plateau after 1.5~m. This is likely caused by increased penumbral blurring at larger distances washing out the dark field signal. Therefore the lung imaging data was collected at 1.5~m. 

\begin{figure}[t!]
\includegraphics{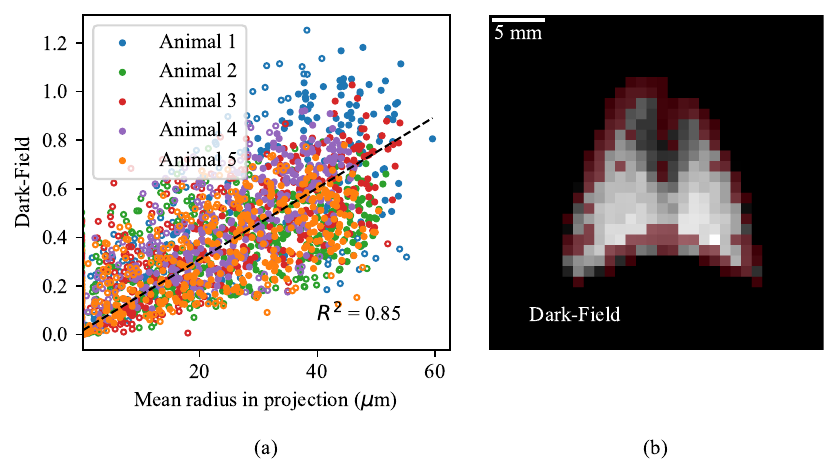}
\centering
\caption{For each rabbit kitten, the dark-field images and alveolar size projections can be compared (as visualised in Figure~\ref{fig:dfr}). Inclusion of all animal data points gives an approximately linear trend as shown in (a). The hollow circles indicate data points where the dark-field value significantly deviates from the linear trendline that has been fitted ($>$50\%). For Animal 1, the origin of these data points is highlighted in red in (b), where we see the data-points are predominantly related to the edges of the lungs. After ignoring the points on the edges of the lungs, we obtain a Pearson's correlation coefficient of $R^2 = 0.85$.}
\label{fig:graphdev}
\end{figure}
\begin{figure}[b!]
\centering
\includegraphics{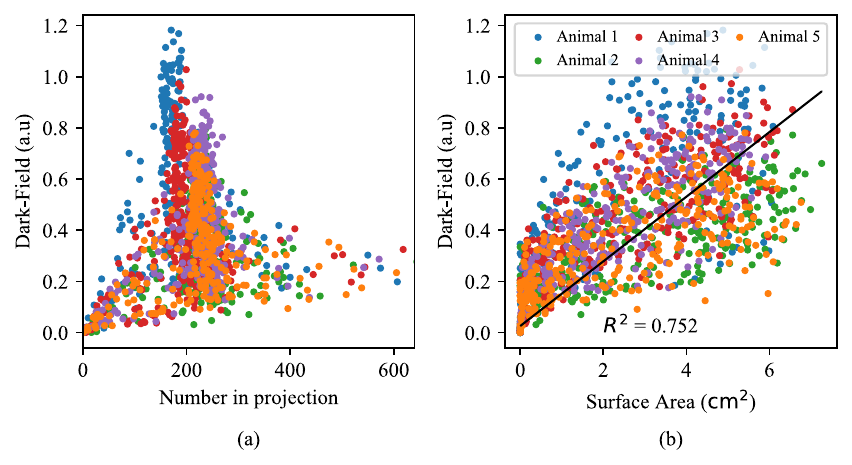}
\caption{Using the granulometry processed volumes (Figures~\ref{fig:dfr} \& \ref{fig:graphdev}), we can calculate the number of alveoli following Equation~\ref{eq:number_form}, this gives (a). Using both the number and size of the alveoli, we can calculate the total surface area, which results in a more linear relationship, shown in (b).}
\label{fig:dfn}
\end{figure}

Using the first projection image of each high-resolution CT acquired (see Section~\ref{sec:experiment}), we retrieved the dark-field signal to compare it with the properties of the projected scatterers (alveoli), such as size, number, and surface area. The alveolar properties were determined by applying 3D granulometry to CT data as follows: (1) applying phase retrieval to each projection before CT reconstruction to boost SNR~\cite{kitchen2017ct}, (2) dividing the volume into cubic regions of interest (from here on these will be referred to as granulometry ROIs); (3) applying a sharpening filter to increase the visibility of the boundary between air and tissue; (4) converting to a binary image of air (ones) and not air (zeros); (5) using a pore size distribution algorithm~\cite{porespy} to determine the radius of the sphere that can fit in each alveolus. The 3D maps of alveolar size were then forward-projected along the optical axis to give a 2D projection of the mean alveolar radius within each row of granulometry ROIs. The chosen granulometry ROI size was 64$\times$64$\times$64 pixels, as this is slightly smaller than regions typically chosen for regional analysis in lung imaging studies~\cite{kitchen2015x}. Choosing a smaller ROI for the granulometry than this would increase the proportion of partial alveoli that would not be counted if they straddle different ROIs for the granulometry, thus reducing the statistical accuracy. We measured each row of granulometry ROIs along the optical axis against the average dark-field measurement in the corresponding 64$\times$64 region. The projected mean radius from granulometry of the CT volume and the re-binned dark-field projection are shown in Figure~\ref{fig:dfr}. 
\begin{figure}[t!]
\centering
\includegraphics{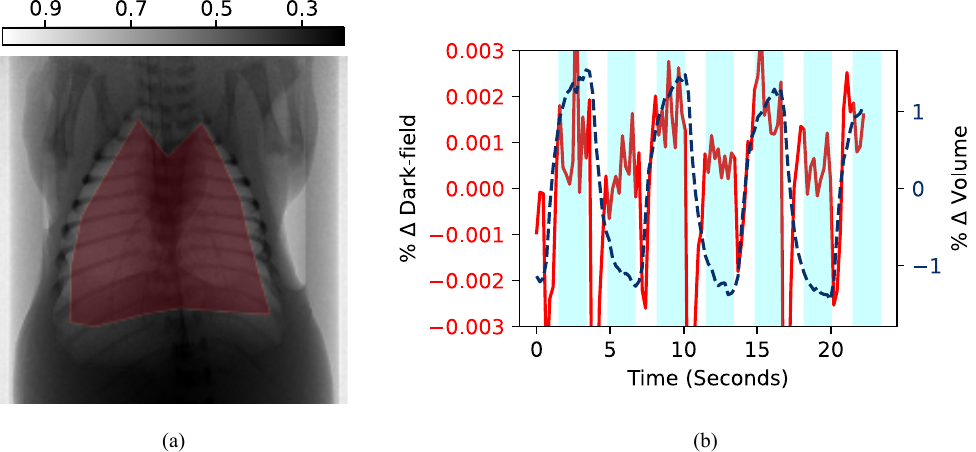 }
\caption{In (a) we have one projection in the recorded ventilation sequence overlaid with a masked region (red) just interior of the lungs. Within this region, the volume change (black) was tracked as a function of time in the ventilation sequence, this is shown in (b) (calculated via~\cite{o2022accurate}). The dark-field was retrieved in each image in the same region. In (b), the \% change in the dark-field and volume are plotted through the ventilation sequence. Due to motion blurring within the PBI projections at points of the start of inflation and deflation, large spikes appear in the dark-field signal. Ignoring these spike artefacts and only comparing the highlighted cyan regions, the correlation between the two measurements is $R=0.52$.}
\label{fig:dynamic_1}
\end{figure}
\begin{figure}[t!]
\centering
\includegraphics{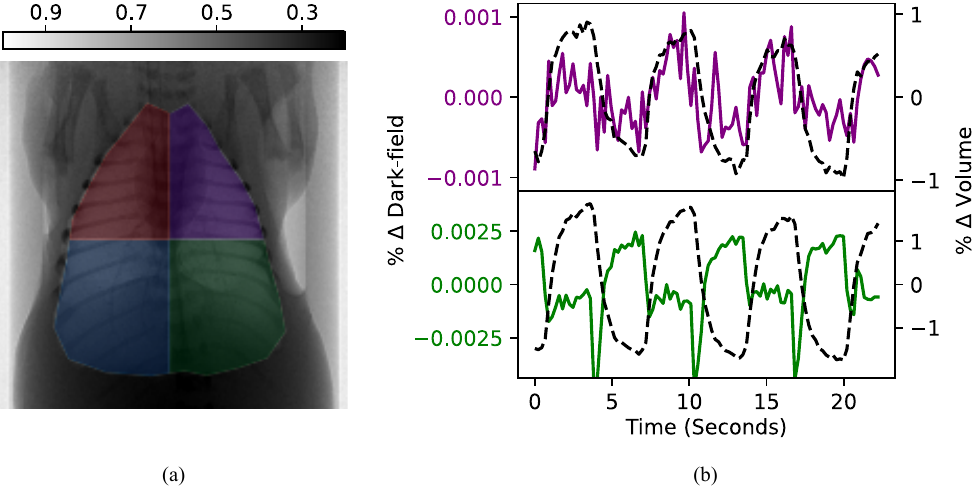}
\caption{In (a), Four quadrants of the lung are shown as coloured overlays on a propagation-based projection from a ventilation sequence. Following a similar procedure in Figure~\ref{fig:dynamic_1}, (b) shows the changes in dark-field compared with changes in the lung air volume for two regions (purple and green). The correlation (Pearson $R$) between the dark-field and air volume changes for the purple and green regions is 0.63 and -0.71 respectively.}
\label{fig:dynamic_2}
\end{figure}

 Figure~\ref{fig:graphdev}(a) shows the dark field signal as a function of the projected mean radius for all five animals.  Each animal data set in Figure~\ref{fig:graphdev}(a) has a similar gradient, showing good agreement between animals with different tissue thicknesses. The data is separated into two categories, shown in the figure with hollow and solid spheres, where the hollow spheres represent data-points that deviate from the given trend line by $> 50 \%$. The anatomical origin of these hollow data-points in `Animal 1' is displayed in Figure~\ref{fig:graphdev}(b) as the highlighted red pixels. This clearly indicates that the dark-field measurements do not scale as consistently with mean alveolar size towards the edges of the lungs, where there are fewer alveoli and a higher proportion of bone. After excluding these data points for each animal, we obtain a Pearson correlation coefficient of $R^2 = 0.85$ for the linear trendline, which shows how the dark-field signal changes with the mean alveolar size.

The projected number of alveoli within a granulometry ROI can be calculated using Equation~\ref{eq:number_form}, upon substituting in the mean alveolar radius for each granulometry ROI, the associated packing fraction of the alveoli, and the thickness of that tissue region. Using the same procedure for measuring the mean radius, we can plot the projected number of alveoli against the dark-field signal, as shown in Figure~\ref{fig:dfn}(a). Here we observe divergent behaviour that was not seen for the PMMA spheres in Section~\ref{sec:spheres} (cf. Figure~\ref{fig:spherethicc}(a) with Figure~\ref{fig:dfn}(a)). Since the total number of alveoli is calculated from the alveolar radius, the inaccuracies at the lung's boundaries (red regions in Figure~\ref{fig:graphdev}(b)) are propagated through, which explains some of the differences between the two studies. Additionally, within each ROI, any scattering from the surrounding tissue, organs, and bone will be mixed with the scattering from the alveoli, creating a non-systematic error in the data set.

\section{Dynamic Dark-field Retrieval}
One of the most significant advantages demonstrated in Sections~\ref{sec:spheres} \&~\ref{sec:lungs} is that only one PBI image is required to retrieve the X-ray dark-field signal. Therefore, we can image the lung over a ventilation sequence (see Section~\ref{sec:experiment}) and retrieve the dark-field signal from each frame. We can, therefore, measure how the dark-field signal changes as the alveoli inflate and deflate. The resultant dark-field can also be averaged over any region of the lungs to improve the SNR of such a measurement. Initially, we chose an ROI over the entire lung (Figure~\ref{fig:dynamic_1}(a)), excluding regions close to the perimeter of the lung in the projection. This exclusion is based on the findings in Section~\ref{sec:lungs}, where dark-field measurements and microstructure properties (such as radii) did not correlate well in these regions. 

The degree of correlation of the dark-field with the air volume in a ventilation sequence is shown in Figure~\ref{fig:dynamic_1}(b). The changes in air volume were calculated from each projection following the lung volume estimation method described in~\cite{o2022accurate}). The Pearson correlation `R' between dark-field signal changes and air volume changes is $R=0.52$ for the highlighted cyan regions, which ignores portions of the ventilation sequence with motion artefacts. These spikes correspond to the steepest gradients in volume at the start of lung inflation and deflation. It is important to highlight the visible difference in the dark-field signal between the different inflation states of the lung. As indicated by the sharp spikes in Figure~\ref{fig:dynamic_1}(b), motion can cause inconsistencies between the retrieved dark-field and the lung air volume. We then chose smaller lung regions with varying degrees of motion and compare them. Figure~\ref{fig:dynamic_2}(a) shows four coloured ROIs (approximately lung quadrants). In Figure~\ref{fig:dynamic_2}(b), we can compare the purple region (top) with the green region (bottom), which respectively show relatively small and large amounts of lung movement. The purple region correlates with changes in dark-field signal and changes in air volume of 0.63, while the green region has a correlation coefficient of -0.71. We again obtain sharp downward spikes attributed to motion for the green region. This leads to an interesting conclusion, where the regions with a smaller degree of movement show a positive correlation between the two measurements, whilst regions with more movement (here, the bottom green section) show dark-field and air-volume changes being anti-correlated. The likely explanation for this anti-correlation is that motion blurring reduces the visibility of phase contrast in the image, leading to a weaker dark-field signal from the blurred tissues boundaries. The difference in behaviour between the purple and green regions could also be of physiological origin of unknown cause. 

\section{Comparison with Previous Techniques}
\begin{figure}[t!]
\centering
\includegraphics[]{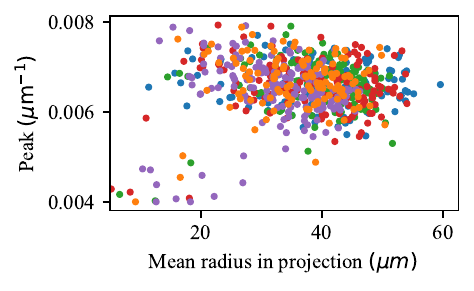}
\caption{Following the methodology from~\cite{kitchen2015x}, we can calculate the dominant frequency of the X-ray speckle for regions of 64$\times$64 in the same PBI projections for the five animals used in Figure~\ref{fig:dfr}. Here, each color corresponds to the same animal as in Figure~\ref{fig:graphdev}}.
\label{fig:bad_speckles}
\end{figure}

As mentioned in Section~\ref{sec:introduction}, regional dominant airway size can be calculated from PBI phase contrast images of lungs~\cite{kitchen2015x}. Kitchen et al. showed that the peak spatial frequency associated with lung speckle reveals the dominant size of the airways that create the speckle. By fitting the speckle power spectrum peak with a Pearson VII function, the centroid of this can then be used to calculate the dominant airway size so, for the case of alveoli in the lung, it allows for the alveolar size to be estimated. However, this method and is also highly sensitive to effects such as motion blurring that reduce the visibility of the speckles. In smaller regions of interest, the fitting procedure is also more difficult, where the power of the speckle pattern in Fourier space is low. In Section~\ref{sec:lungs}, we have demonstrated a strong correlation between the average dark-field signal and the alveolar size (Figure~\ref{fig:graphdev}). Using the same data set from Section~\ref{sec:lungs}, we then compared how well Kitchen et al.'s method performs against the results obtained herein. This is shown in Figure~\ref{fig:bad_speckles}, where we see a lower correlation between the dominant spatial frequency and the mean radius in projection using the speckle measurements. Quantitatively, we obtain an $R^2$ of just $0.27$ for these measurements. This correlation appears much weaker than that found in the original paper by Kitchen et al., potentially due to the reduced coherence of IMBL compared to beamline BL20B2 at the SPring-8 synchrotron, where the data for that study was collected. Another major difference is the small region of interest used here of 64$\times$64 pixels compared to their studies using regions no smaller than 128$\times$128 pixels. For both these reasons, the Fourier space peak here will be weaker, and the subsequent analysis of Kitchen et al.~\cite{kitchen2015x} may not perform as well on this dataset. Conversely, it is encouraging how well the dark-field algorithm of Gureyev et al.~\cite{gureyev_dark-field_2020}  has worked here with the small regions-of-interest and modest spatial coherence.

\section{Discussion}
\label{sec:discuss}
As the dark-field signal retrieved in this study is proportional to both the scatterer's number and size, we cannot directly take the retrieved dark-field signal and infer the measurement of interest. For example, one may wish to determine the size of the alveoli in the lungs to assess the damage caused by emphysema (a form of COPD). However, the number of alveoli in projection will have a confounding influence. This problem is also present in other interferometric dark-field retrieval methods that calculate a dark-field coefficient~\cite{bech2010quantitative, urban2022qualitative}. In these methods, the tissue thickness must also first be estimated. 

The dark-field algorithm used herein (as discussed in Section~\ref{sec:alg}) seeks to solve for the dark-field signal from the media interfaces. However, any resolvable intensity variations due to diffusive dark-field will also be present. Indeed, given that the retrieved signal is influenced by both the size and the number of the scattering objects, it suggests that we are observing a blend of refractive and diffractive effects.

The inability to isolate measurements of scatterer size or number is a disadvantage. However, a combination of these measurements (i.e. the total alveolar surface area) is an important quantity in lung imaging. From the results, we can, in practice, use the dark-field signal to quantify surface area by using a calibration curve (e.g. Figure~\ref{fig:dfn}(b)). This is important, since the total lung surface area is a primary driver in the rate of gas exchange in the lungs ~\cite{hsia2016lung}.  Other dark-field approaches based on PBI, such as those using the Fokker–Planck equation~\cite{morgan_applying_2019, paganin2019x}, have suggested that their retrieved tomographic dark-field signal is related to the surface area to volume ratio~\cite{leatham2023x, leatham2024x}, which is in agreement with our results. 

We have shown how lung surface area can be estimated from a single image, and from this we have developed a technique that can assess alveoli microstructure, also using only a single image. This enables higher temporal resolution than high-resolution CT and utilises fewer specialised optics than other dark-field imaging methods. The technique has been successfully applied using a laboratory microfocus X-ray source (Section~\ref{sec:spheres}) for static samples and using synchrotron radiation where a higher flux was required to image respiratory motion in real-time (Section~\ref{sec:lungs}). 

In this work, we have investigated the difference in the dark field signal retrieved from edge scattering compared to that retrieved from the blurring of a reference pattern, which is better described as an intensity diffusion \cite{paganin2019x, morgan_applying_2019}. We have shown a dependence on both scatterer size and number, which cannot be determined using ABI and single grid techniques, where the retrieved signal is shown to be dependent only on the number of interfaces. Another key difference with this method is that sufficient contrast resolution of the phase and scattering effects is required to obtain dark-field retrieval. Hence, a sufficiently large propagation distance and high spatial coherence is required. However, as both the edge scattering and diffusion due to unresolved microstructure occur in lungs, these measurements are able to give similar microstructure information and could be used complementarily.   
Further investigations will aim to explore the feasibility of imaging larger animals with significantly more attenuating tissue. In addition, wider adoption will be helped by transitioning the dynamic technique to laboratory-based sources, aided by the use of higher power X-ray sources, such as liquid metal jet sources~\cite{larsson201124}) and more efficient detectors (photon-counting~\cite{donath2023eiger2}, or non-photon counting~\cite{o2020photon}). 

\section{Conclusion}
We have investigated the relationship between a dark-field signal retrieved from a single propagation-based exposure (following Gureyev et al. 2020) and the microstructure properties of both dark-field-generating PMMA spheres and alveoli within rabbit kitten lungs. We found that the retrieved signal varies with the projected size and the number of alveoli along the optical axis. Notably, the retrieved dark-field signal is correlated ($R^2 = 0.752$) with the projected surface area of the microstructures. For diagnostic imaging, this gives an important measurement of lung health, including structure and function, which can be estimated from a single propagation-based phase-contrast X-ray image. Our experiments were performed on small mammals, meaning further research should focus on the barriers towards clinical implementation. 

\section*{Funding}
This work was funded in part by the Australian Research Council’s Discovery Grant
DP170103678, the Future Fellowship Schemes under Grant
FT160100454 and Grant FT180100374, the National Health
and Medical Research Council (NHMRC) Project under Grant 1102564, the German Excellence Initiative, the European Union
Seventh Framework Program under Grant 291763. This work is supported by the Victorian Government’s Operational Infrastructure Support Program and the Australian Government under
Grant AS/IA173/14218. Erin V. McGillick was supported by a NHMRC Peter Doherty Biomedical Early Career Fellowship (APP1138049). Dylan W. O'Connell was supported by the Faculty of Science Dean’s Postgraduate Research Scholarship awarded by the School of Physics and Astronomy at Monash University and the Research Training Program (RTP) Scholarship.

\bibliographystyle{IEEEtran}  
\bibliography{output}

\end{document}